\documentclass{article}
\usepackage{spconf,amsmath,graphicx}
\usepackage{amssymb}
\usepackage{color}
\usepackage{caption}
\usepackage{subcaption}
\usepackage{float}

\def\0{{\mathbf 0}}
\def\1{{\mathbf 1}}

\def\x{{\mathbf x}}

\def\A{{\mathbf A}}

\def\D{{\mathbf D}}

\def\I{{\mathbf I}}

\def\P{{\mathbf P}}

\def\W{{\mathbf W}}
\def\X{{\mathbf X}}

\def\cF{{\mathcal F}}
\def\cG{{\mathcal G}}

\def\bPhi{{\boldsymbol \Phi}}

\title{Volumetric 3D Point Cloud Attribute Compression: \\ Learned polynomial bilateral filter for prediction}
\name{
    Tam Thuc Do$^\dag$,
    Philip A. Chou, 
    Gene Cheung$^\dag$
}

\address{
  $^\dag$York University, Toronto, Canada
}

\begin{document}
\ninept
\maketitle
\begin{abstract}
We extend a previous study on 3D point cloud attribute compression scheme that uses a volumetric approach: given a target volumetric attribute function $f : \mathbb{R}^3 \mapsto \mathbb{R}$, we quantize and encode parameters $\theta$ that characterize $f$ at the encoder, for reconstruction $f_{\hat{\theta}}(\x)$ at known 3D points $\x$ at the decoder. 
Specifically, parameters $\theta$ are quantized coefficients of B-spline basis vectors $\bPhi_l$ (for order $p \geq 2$) that span the function space $\cF_l^{(p)}$ at a particular resolution $l$, which are coded from coarse to fine resolutions for scalability.
In this work, we focus on the prediction of finer-grained coefficients given coarser-grained ones by learning parameters of a polynomial bilateral filter (PBF) from data. 
PBF is a pseudo-linear filter that is signal-dependent with a graph spectral interpretation common in the graph signal processing (GSP) field.
We demonstrate PBF's predictive performance over a linear predictor inspired by MPEG standardization over a wide range of point cloud datasets. 
% Extending a previous work Region Adaptive Hierarchical Transform (RAHT) that employs piecewise constant functions to span a nested sequence of function spaces, we propose a feedforward linear network that implements higher-order B-spline bases spanning function spaces without eigen-decomposition. 
% Feedforward network architecture means that the system is amenable to end-to-end neural learning. 
% The key to our network is space-varying  convolution, similar to a graph operator, whose weights are computed from the known 3D geometry for normalization.
% We show that the number of layers in the normalization at the encoder is equivalent to the number of terms in a matrix inverse Taylor series.
% Experimental results on real-world 3D point clouds show up to 2-3 dB gain over RAHT in energy compaction and 20-30\% bitrate reduction. 
\end{abstract}
\vspace*{-0.5ex}
\section{Introduction}
\label{sec:intro}

We study lossy compression of 3D point cloud attributes from a volumetric function approach.
Specifically, assuming point cloud \textit{geometry} is first encoded, we encode point cloud \textit{attributes}, given that encoded geometry is known both at encoder and decoder---this is the dominant approach both in research \nocite{ZhangFL:14,ThanouCF:16,QueirozC:16,QueirozC:17b,CohenTV:16,ChouKK:20,KrivokucaCK:20}\cite{ZhangFL:14}--\cite{KrivokucaCK:20} and in MPEG geometry-based point cloud compression (G-PCC) standard \nocite{SchwarzEtAl:18,GraziosiEtAl:20,JangEtAl:19,GPCC:20}\cite{SchwarzEtAl:18}--\cite{GPCC:20}. 
Mathematically, given known 3D locations $\x_i \in \mathbb{R}^3$ both at the encoder and decoder, we encode quantized parameters $\hat{\theta}$ for a target volumentric attribute function $f: \mathbb{R}^3 \mapsto \mathbb{R}$ from coarse to fine resolutions for scalability, so that it can be evaluated as $f_{\hat{\theta}}(\x_i)$ at $\x_i$ at the decoder for signal reconstruction. 
\cite{QueirozC:16,SandriCKQ:19} proposed such a framework using volumetric B-spline basis functions $\bPhi_l$ of order $p=1$ that span a nested sequence of function spaces $\cF_{l_0}^{(p)} \subseteq  \cdots \subseteq \cF_L^{(p)} $---called \textit{Region Adaptive Hierarchical Transform} (RAHT(1))---and now forms the core in MPEG G-PCC \cite{ChouKK:20}. 
Recently, \cite{do2023volumetric} extended this framework to B-spline basis functions of order $p \geq 2$ (RAHT($p$)) and demonstrated state-of-the-art (SOTA) coding performance. 
Moreover, the feedforward network obtained by unrolling a finite-term Taylor's series of a matrix inverse to compute orthonormalized RAHT($p$) coefficients $\bar{F}_l^* = (\bPhi_l^\top \bPhi_l)^{-1/2} \bPhi_l^\top f$ is amenable to end-to-end data-driven parameter tuning. 
The goal of this paper is to further improve performance by designing a predictor for finer-grained unnormalized coefficients $F_{l+1}^*$ given coarser-grained coefficients $F_l^*$, and training. 

In particular, stemming from the seminal bilateral filter \cite{tomasi98} in image denoising, we propose a \textit{polynomial bilateral filter} (PBF) that is a signal-dependent \textit{pseudo-linear} filter for finer-grained coefficient prediction. 
PBF has been shown in \cite{antonio2013polybf} to possess a graph spectral low-pass filter interpretation that is common in the \textit{graph signal processing} (GSP) liteature \cite{ortega18ieee,cheung18}.
Leveraging the feedforward charcteristic of the developed network \cite{do2023volumetric}, we train the coefficients of PBF and other network parameters in a data-driven manner for optimal prediction. 
Experimental results show that PBF's predictive performance outperforms a linear predictor inspired by MPEG standardization over a wide range of point cloud datasets.

The outline of the paper is as follows.
We first overview the coding framework in Section\;\ref{sec:framework}.
We describe our proposed predictor and prediction residual coding in Section\;\ref{sec:implement}.
Finally, we present experimental results and conclusion in Section\;\ref{sec:results} and \ref{sec:conclude}, respectively.

\vspace*{-0.5ex}
\section{Coding Framework}
\label{sec:framework}

\begin{figure}[t]
    \centering
    % trim=left bottom right top, clip
    % \begin{subfigure}{0.2\textwidth}
    %     \centering
    %     \includegraphics[width=\linewidth,trim=7.0in 5.0in 7.0in 2.5in,clip]{ICASSP24/Figures/Original00.png}
    %     \caption{}
    % \end{subfigure}
    \begin{subfigure}{0.15\textwidth}
        \centering
        \includegraphics[width=\linewidth,trim=9.5in 8in 8.3in 3.8in,clip]{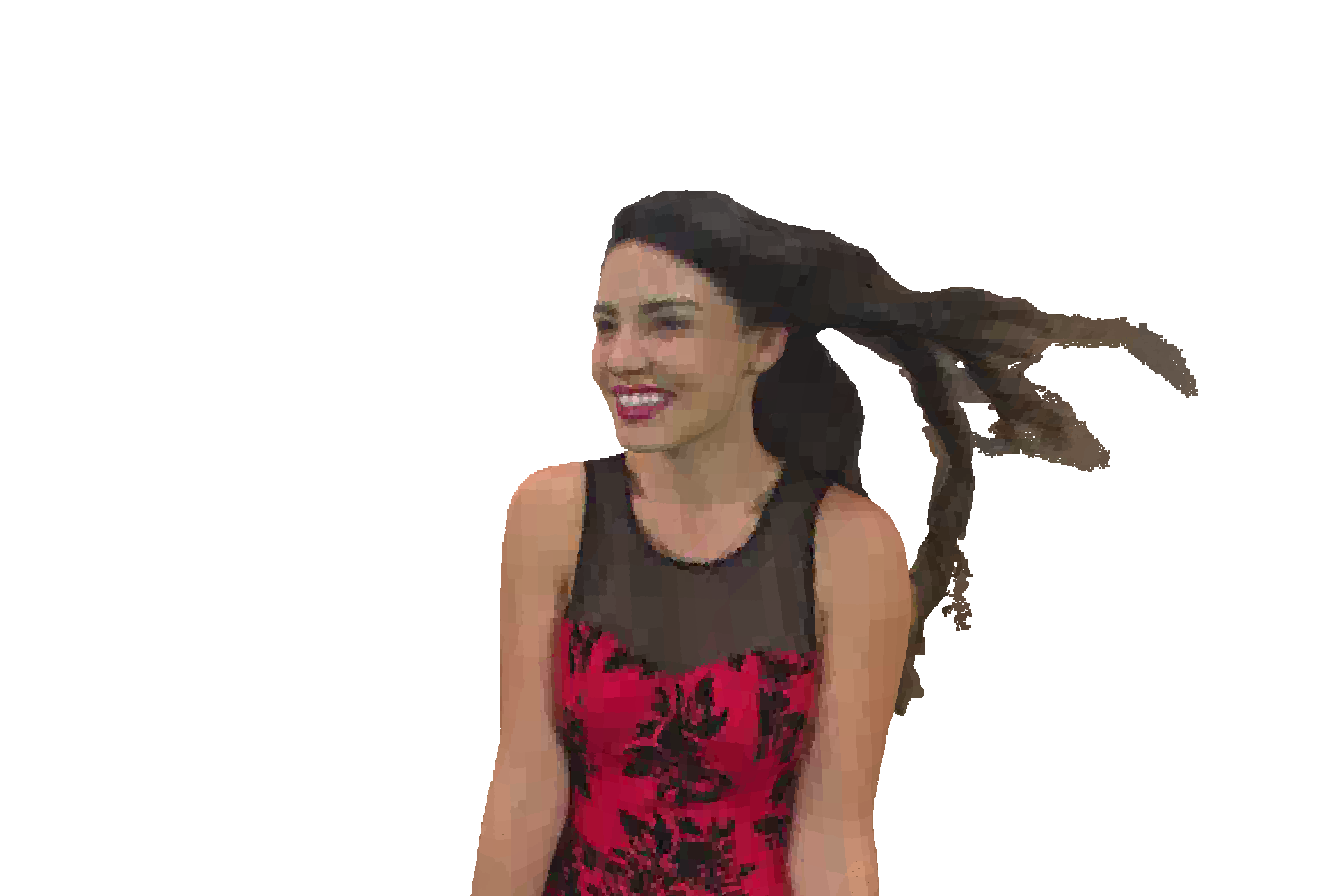}
        \caption{}
    \end{subfigure}
    \begin{subfigure}{0.15\textwidth}
        \centering
        \includegraphics[width=\linewidth,trim=9.5in 8in 8.3in 3.8in,clip]{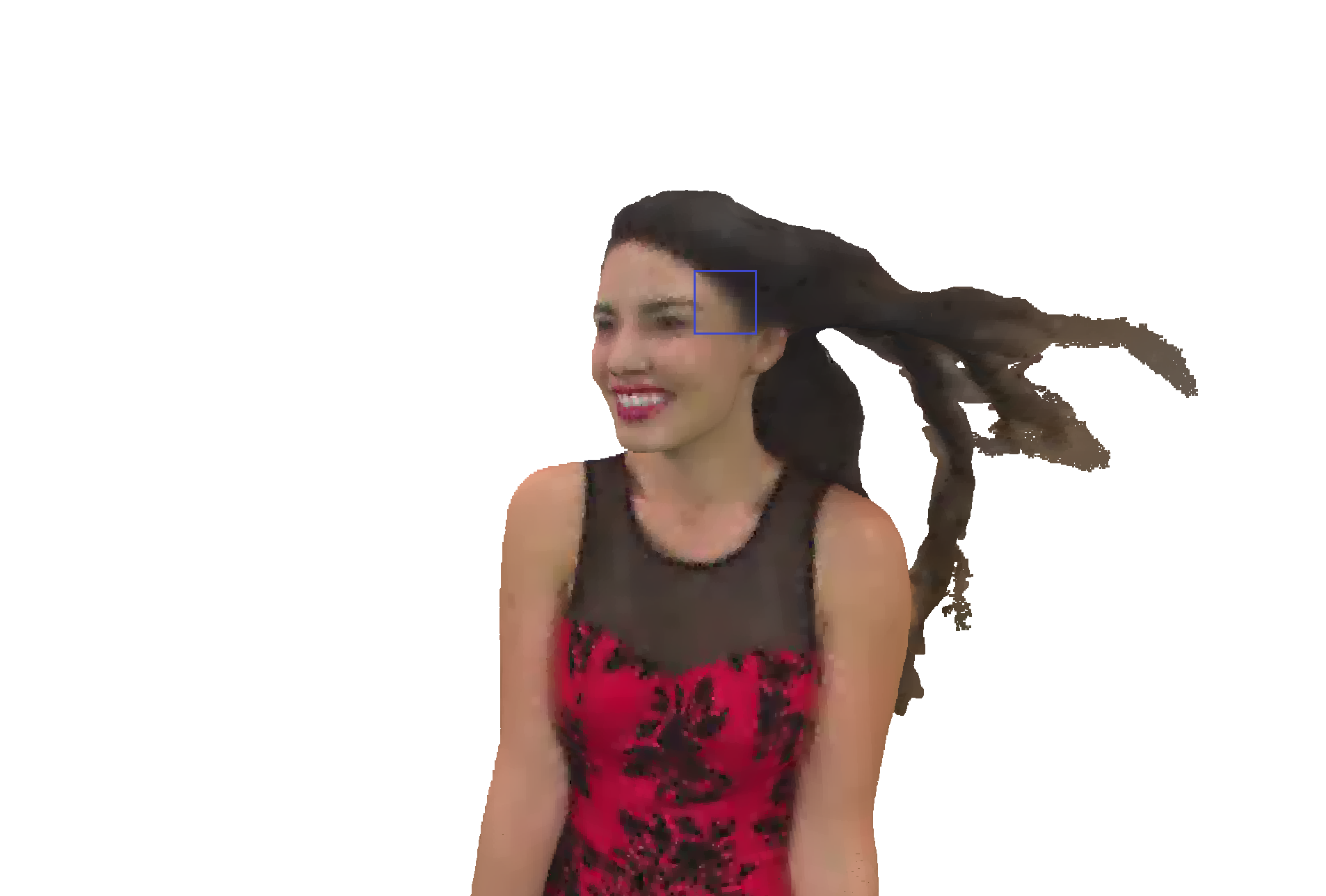}
        \caption{}
    \end{subfigure}
    \begin{subfigure}{0.15\textwidth}
        \centering
        \includegraphics[width=\linewidth,trim=9.5in 8in 8.3in 3.8in,clip]{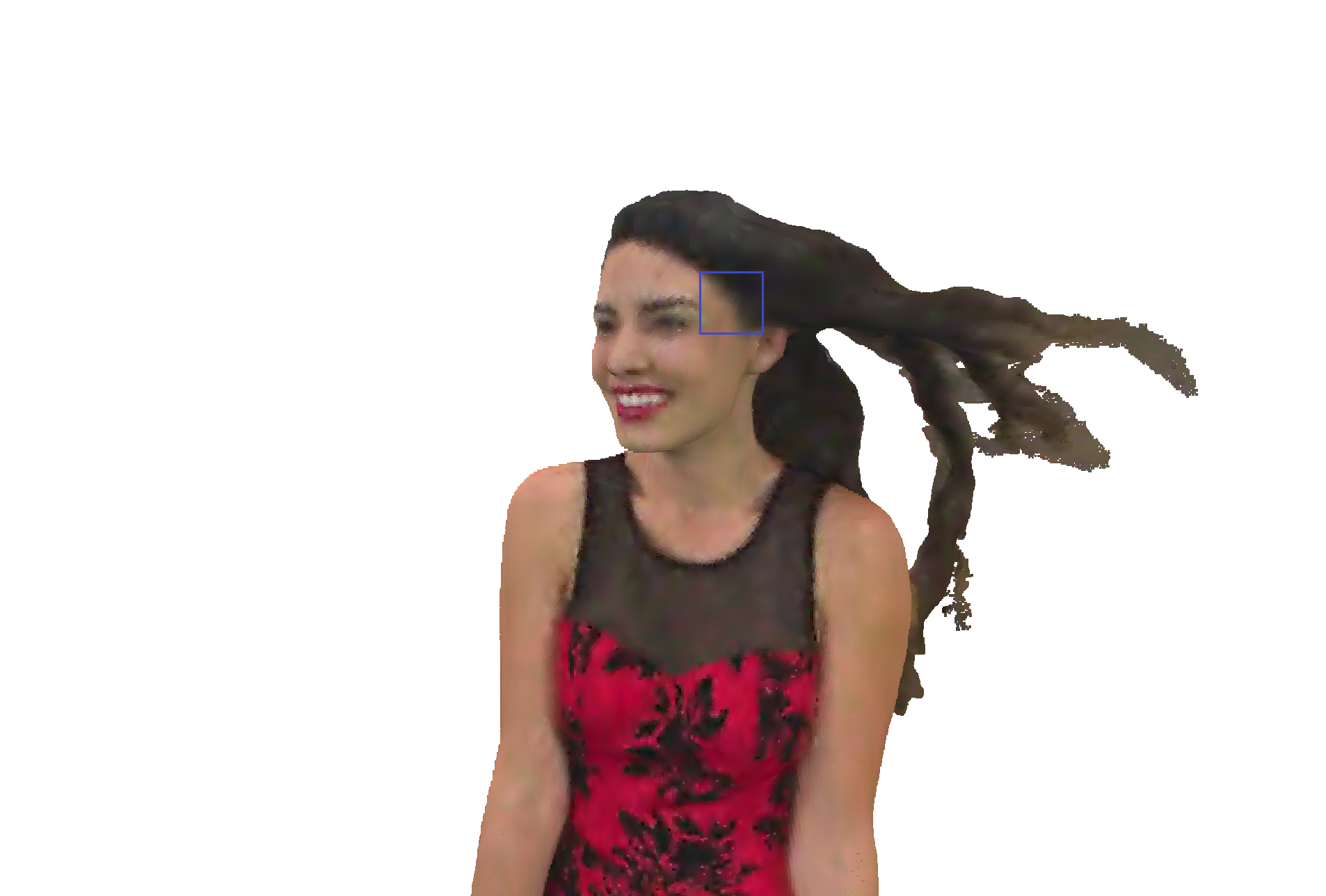}
        \caption{}
    \end{subfigure}
    \vspace{-0.05in}
    \caption{Point cloud encoded at $0.325$bpp: (a) Naive RAHT($p=1$); (b) RAHT($p=1$) + linear; (c) Our model RAHT($p=2$) + PBF.}
    \label{fig:reconstructed_PCs}
\end{figure}

Denote by $f:\mathbb{R}^3\mapsto\mathbb{R}$ a volumetric function defining an attribute field over $\mathbb{R}^3$; the extension to vector attributes is straightforward.
Then, a point cloud can be seen as samples of $f$, with attributes $y_i=f(\mathbf{x}_i)$, where $\x_i \in \mathbb{R}^3$ is sample $i$'s 3D position. 
In \cite{do2023volumetric}, RAHT($p$) codes the attributes given the positions by projecting $f$ onto a nested sequence of function spaces $\mathcal{F}_{l_0}^{(p)}\subseteq\cdots\subseteq\mathcal{F}_L^{(p)}$. Each function space $\mathcal{F}_l^{(p)}$ is a parametric family of functions spanned by a set of basis functions $\{\phi_{l,\mathbf{n}}^{(p)}:\mathbf{n}\in\mathcal{N}_l\}$, namely the volumetric B-spline basis functions $\phi_{l,\mathbf{n}}^{(p)}(\mathbf{x})=\phi_{0,\mathbf{0}}^{(p)}(2^l\mathbf{x}-\mathbf{n})$ of order $p$ and scale $l$ for offsets $\mathbf{n}\in\mathcal{N}_l\subset\mathbb{Z}^3$.
The results are the projections $f_l^*$ of $f$ on $\mathcal{F}_l$ and the residual functions between consecutive projections, defined as $g_l^* = f_{l+1}^* - f_l^*$. 
Denote by $\mathcal{G}_l$ the orthogonal complement subspace to $\mathcal{F}_l$ in $\mathcal{F}_{l+1}$. 
Then, we can write
\vspace{-1.3ex}
\begin{eqnarray}
\cF_L & = &\cF_{l_0} \oplus \cG_{l_0+1} \oplus \cdots \oplus \cG_l \oplus \cdots \cG_{L-1}   \;\;\; \mbox{and}
\label{eq:sum_of_subspaces} \\
f_L^* & = & f_{l_0}^* + g_{l_0+1}^* + \cdots + g_l^* + \cdots + g_{L-1}^* .
\vspace{-2.4ex}
\label{eqn:fg0_gL}
\end{eqnarray}
We assume that the resolution $L$ is high enough such that $f_L^*(\mathbf{x}_i)=f(\mathbf{x}_i)$.
We then code $f$ by coding $f_{l_0}^*$, $g_{l_0+1}^*,\ldots,g_{L-1}^*$.  
These functions are represented as coefficients in a basis, which are quantized and entropy coded.  
\cite{do2023volumetric} showed how to compute these coefficients efficiently. 
We describe this next.

\subsection{Low-pass coefficients}
\label{sec:low_pass_coefs}
Denote by $\bPhi_l=[\phi_{l,\mathbf{n}}]$ the row-vector of basis functions at level $l$, and by $\bPhi_l^\top=[\langle\phi_{l,\mathbf{n}},\cdot\rangle]$ the column-vector of functionals that are inner products with the basis functions at level $l$. 
Then $\bPhi_l^\top f=[\langle\phi_{l,\mathbf{n}},f\rangle]$ is the column-vector of inner products of the basis functions with $f$, and $\bPhi_l^\top\bPhi_l$ is the matrix of inner products of the basis functions with themselves, namely the Gram matrix.  Here, the inner product is given by $\langle g,f\rangle=\sum_ig(\mathbf{x}_i)f(\mathbf{x}_i)$.

Further, denote by $f_l^*$ the projection of the function $f$ onto the subspace $\mathcal{F}_l$, and by $F_l^*$ its coefficients in the basis $[\phi_{l,\mathbf{n}}]$, i.e., $f_l^*=\bPhi_{l,\mathbf{n}}F_l^*$.  The coefficients $F_l^*$, which we call {\em low-pass coefficients}, can be determined by solving the normal equations:
\vspace{-1ex}
\begin{equation} \label{eq:normal_equation_F}
F_{l}^* = (\bPhi_{l}^{\top} \bPhi_{l})^{-1} \bPhi_{l}^{\top} f .
\vspace{-1ex}
\end{equation}

Assume that there is a sufficiently fine resolution $L$ such that the $i$-th basis function $\phi_{L,\mathbf{n}_i}$ is 1 on $\mathbf{x}_i$ and 0 on $\mathbf{x}_j$ for $j\neq i$, namely $\phi_{L,\mathbf{n}_i}(\mathbf{x}_j)=\delta_{ij}$.  
Then, we have $\bPhi_L^\top \bPhi_L = \I$.
Next, denote by $\A_{l}=[a_{ij}^l]$ a $N_{l} \times N_{l + 1}$ matrix whose $i$th row expresses the $i$-th basis function $\phi_{l,\mathbf{n}_i}$ of $\mathcal{F}_l$ in terms of the basis functions $\phi_{l+1,\mathbf{m}_j}$ of $\mathcal{F}_{l+1}$, namely
\begin{equation}
\bPhi_l = \bPhi_{l+1} \A_l^\top .
\label{eqn:Phi_ell_in_terms_of_Phi_ell1}
\end{equation}
% then projecting funtion $f$ on to subspace $\mathcal{F}_l$ would involve solving for the normal equation, namely
% \vspace{-1ex}
% \begin{equation} \label{eq:normal_equation_F}
% f_l^* = \bPhi_{l} F_{l}^* = \bPhi_{l} (\bPhi_{l}^{\top} \bPhi_{l})^{-1} \bPhi_{l}^{\top} f .
% \vspace{-1ex}
% \end{equation}
% where $F_{l}^*$ is the coefficients that represent $f_l^*$ in the basis $\bPhi_{l}$.
Defining $\mathbf{d}_{ij} \triangleq \mathbf{m}_j-2\mathbf{n}_i$, 
$\A_{l}$ may be parameterized by a vector of size $3\times 3 \times 3 = 27$ as
\begin{equation}
a_{ij}^l = \left\{\begin{array}{ll}
w_{\mathbf{d}_{ij}} & \mbox{if}~~ \mathbf{d}_{ij} \in \{-1, 0, 1\}^3 \\ 
0 & \mbox{otherwise}
\end{array}\right .
\label{eq:RAHTp}
% \vspace*{-2ex}
\end{equation}
Thus, $\mathbf{A}_l$ is a space-invariant but {\em sparse} convolutional matrix with $2\times$ spatial downsampling, and $\mathbf{A}_l^\top$ is the corresponding transpose convolutional matrix with $2\times$ upsampling.
% \red{not sure what you mean.} \blue{Thuc: the scale of the basis function are multiples of $2^{l}$ so $\mathbf{A}_l^\top$ work like a upsampling operation}.  
Using these, the Gram matrix at level $l$ can be computed from the Gram matrix at level $l+1$:
% \vspace*{-1ex}
\begin{eqnarray}
\bPhi_l^\top \bPhi_l & = & \A_l\bPhi_{l+1}^\top \bPhi_{l+1} \A_l^\top .
% \vspace*{-1ex}
\label{eqn:PhiTPhi}
\end{eqnarray}
The matrix $\bPhi_l^\top \bPhi_l $ can also be interpreted as a sparse convolutional matrix, though space-varying, corresponding to the edge weights of a graph with 27-neighbor connectivity (including self-loops).  Hence it can be represented as a $N_l\times 27$ tensor, here denoted by $[[\bPhi_l^\top\bPhi_l]]$.  
This allows us to define a \textit{geometric attention} operator $\mathbf{\Gamma}_l$ as
\vspace*{-1ex}
\begin{eqnarray}
\mathbf{\Gamma}_l[[\bPhi_{l+1}^\top \bPhi_{l+1}]] & = & [[\A_l\bPhi_{l+1}^\top \bPhi_{l+1} \A_l^\top]] .
\label{eqn:GeoAttPhiTPhi}
% \vspace*{-1ex}
\end{eqnarray}
It is possible to see that $\mathbf{\Gamma}_l$ is a sparse convolution matrix with a space-invariant $3^3 27^2$ kernel derived from the $3^3$ kernel of $\A_{l}$.

Thus, beginning at the highest level of detail $L$, where 
% $\phi_{L,\mathbf{n}_i}(\mathbf{x}_j)=\delta_{ij}$ and 
$\bPhi_L^\top\bPhi_L=\I$, $\bPhi_L^\top f=[y_i]$, and $F_L^*=(\bPhi_L^\top\bPhi_L)^{-1}\bPhi_L^\top f=[y_i]$, for each $l<L$ the low-pass coefficients can be expressed recursively,
% we then solve the normal equation to get the projection $f_l^*$ at each level \cite{do2023volumetric, ChouKK:20}. Firstly,  $F_L^*=(\bPhi_L^\top\bPhi_L)^{-1}\bPhi_L^\top f=[y_i]$, then moving to $l<L$, we have
\vspace*{-1ex}
\begin{align}
F_l^* & = (\bPhi_l^\top \bPhi_l)^{-1} \bPhi_l^\top f \\
& = (\bPhi_l^\top \bPhi_l)^{-1} \bPhi_l^\top f_{l+1}^* \\
& = (\bPhi_l^\top \bPhi_l)^{-1} \bPhi_l^\top \bPhi_{l+1} F_{l+1}^* 
\label{eqn:Fl} \\
& = (\bPhi_l^\top \bPhi_l)^{-1} \A_l(\bPhi_{l+1}^\top \bPhi_{l+1}) F_{l+1}^*.
\label{eqn:F_ell_star_from_F_ell1_star}
% \vspace*{-1ex}
\end{align}
If we let $\tilde{F}_l^* = (\bPhi_{l}^\top \bPhi_{l}) F_{l}^*$, which we call {\em un-normalized} low-pass coefficients, we can easily calculate
$\tilde{F}_l^*$ by an ordinary sparse convolution for all levels $l$ using \eqref{eqn:analysis_lowpass}, then calculate $F_l^*$ using \eqref{eqn:normalization}:
\vspace{-1ex}
\begin{align}
    \tilde{F}_l^* &= \A_l \tilde{F}_{l+1}^* \label{eqn:analysis_lowpass}\\
    F_l^* &= (\bPhi_{l}^\top \bPhi_{l})^{-1} \tilde{F}_l^* \label{eqn:normalization}.
    \vspace{-3ex}
\end{align}
To compute \eqref{eqn:normalization}, we implement the operator $\mathbf{X}^{-1}$ as a $P$-layer feedforward network with coefficients initially derived from the $P$-th order Taylor expansion of $x^{-1}$. 
Later on, these Taylor coefficients may be further optimized in a training process introduced in Section\;\ref{sec:results}.

\vspace*{-1.5ex}
\section{Prediction \& Residual Coding}
\label{sec:implement}

\begin{figure}[t]
    \centering
    \includegraphics[%width=1\linewidth,
    height=0.3\textheight, trim=1.0in 0.125in 5.125in 0.25in,clip]{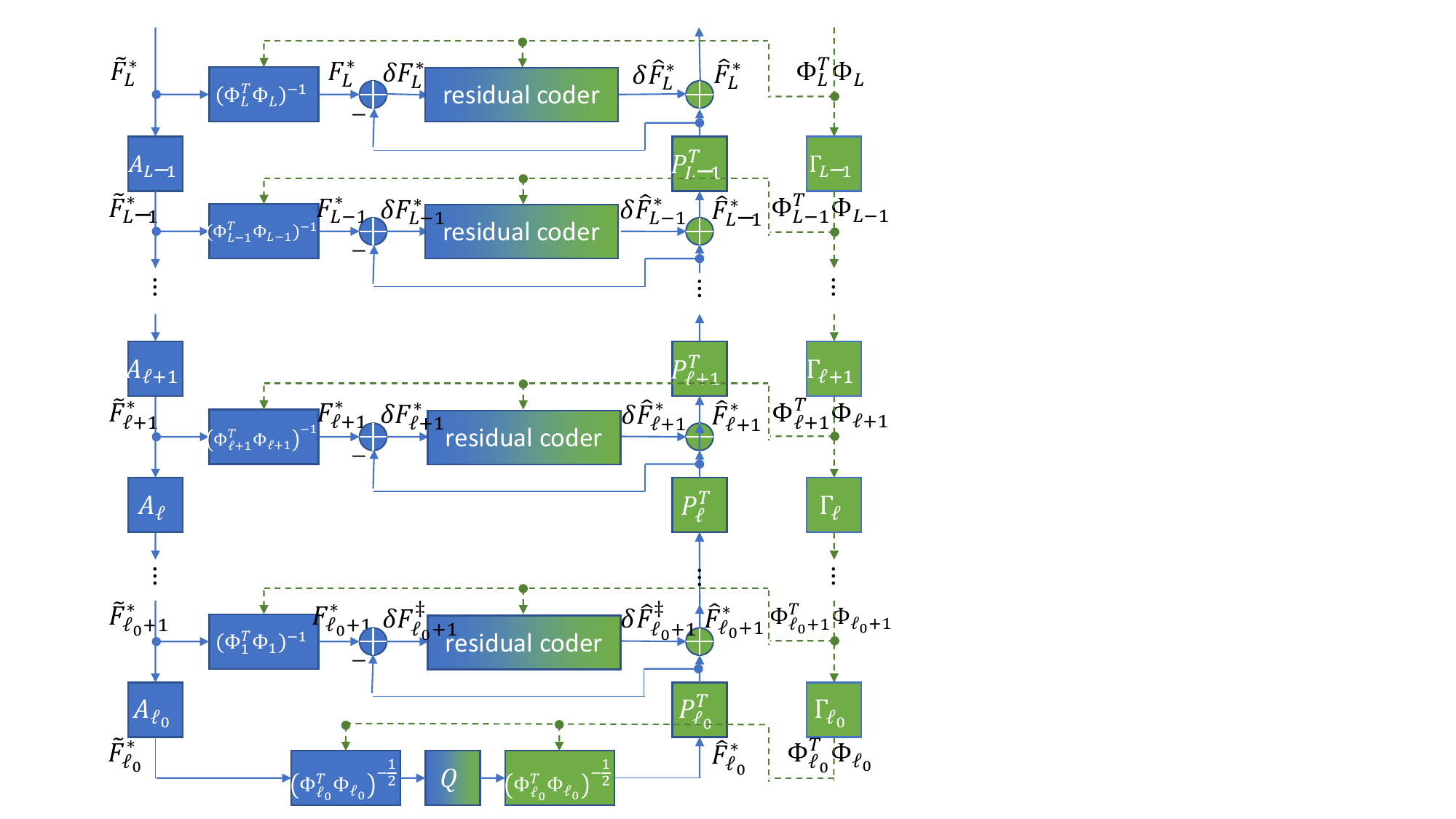}
    \vspace{-0.1in}
    \caption{Multilayer feedforward network implementing point cloud attribute encoder (blue and green) and decoder (green).}
    \label{fig:feedforward_network_block_diagram}
    \vspace*{-3ex}
    
\end{figure}
\vspace{-0.05in}

\subsection{Residual Coding} \label{sec:residual_coding}
As introduced in \cite{do2023volumetric}, it is possible to represent the residual function $g_l^*$ in the basis $\bPhi_{l+1}$. In this basis, the coefficients for $g_l^*=f_{l+1}^*-f_l^*=\bPhi_{l+1}F_{l+1}^*-\bPhi_l F_l^*$ are given by (\ref{eq:normal_equation_F}) and (\ref{eqn:Phi_ell_in_terms_of_Phi_ell1}) as
\vspace{-1ex}
\begin{equation}
    (\bPhi_{l+1}^\top\bPhi_{l+1})^{-1}\bPhi_{l+1}^\top g_l^* = F_{l+1}^*-\A_l^\top F_l^* \stackrel{\Delta}{=} \delta F_{l+1}^* .
    \label{eqn:delta_F}
\vspace{-0.5ex}
\end{equation}
We call these {\em residual coefficients}.

In the present work, we reduce the magnitude of these residual coefficients by generalizing the matrix $\A_l^\top$ from a simple $2\times$ upsampler followed by linear interpolation to a more general $2\times$ upsampling prediction matrix $\P_l^\top$, similar to a super-resolution module. 
The inputs to this module are the coefficients that represent $f_l^*$ at resolution $l$, and the outputs are the prediction of coefficients that represent $f_{l+1}^*$ at level $l+1$. 
Thus, \eqref{eqn:delta_F} becomes
\begin{equation}\label{eq:predict_module}
    \delta F_{l+1}^{\ddagger} \triangleq  F_{l+1}^* - \P_l^\top F_l^* ,
\vspace{-0.5ex}
\end{equation}
where $\delta F_{l+1}^{\ddagger}$ are residual coefficients in the basis $\bPhi_{l+1}$ as above, generally smaller in magnitude than $\delta F_{l+1}^*$.  Like $\A_l^\top$, $\P_l^\top$ is an up-sampling matrix of size $N_{l+1}\times N_{l}$. However, while $\A_l^\top$ is a simple transpose convolution operator with a space-invariant $3\times3\times3$ kernel, $\P_l^\top$ may be the composition of many convolution operators with $3\times3\times3$ kernels, which may be space-varying, may be non-linear, and may involve additional up and down sampling.

As done in \cite{do2023volumetric}, instead of quantizing the coefficients $F_l^*$ or $\delta F_{l+1}^{\ddagger}$ directly, we \textit{orthonormalize} them first as
\vspace{-1ex}
\begin{align}
    \bar{F}_{l}^* &
    % = (\bPhi_{l}^{\top} \bPhi_{l})^{-\frac{1}{2}} \bPhi_{l}^{\top} f
    = (\bPhi_{l}^{\top} \bPhi_{l})^{\frac{1}{2}} F_l^* \label{eqn:orthor_normal_equation_F} \;\;\;\mbox{and} \\
    \delta \overline{F}_{l+1}^{\ddagger}  &
    % = (\bPhi_{l+1}^{\top} \bPhi_{l+1})^{-\frac{1}{2}} \bPhi_{l+1}^{\top} g_{l}^*
    = (\bPhi_{l+1}^{\top} \bPhi_{l+1})^{\frac{1}{2}} \delta F_{l+1}^{\ddagger} .
\label{eq:orthor_normal_equation_G}
\end{align}
To compute \eqref{eqn:orthor_normal_equation_F} and \eqref{eq:orthor_normal_equation_G}, we implement the operator $\X^{-\frac{1}{2}}$ as a $P$-layer feedforward network with coefficients initially derived from the $P$th order Taylor expansion of $x^{-\frac{1}{2}}$. Thus, the orthonormalized coefficients will correspond to the orthonormal basis functions $\bar{\bPhi}_l = \bPhi_l (\bPhi_l^\top \bPhi_l)^{-\frac{1}{2}}$, which are then quantized for entropy coding.

\subsection{Locally Linear Prediction} \label{sec:locally_linear_enhancement}

Inspired by MPEG G-PCC \cite{SchwarzEtAl:18, Wang2023predGPCC, GPCC:21}, we first introduce a \textbf{locally linear weighted average} (LLWA) operator as the basic enhancement operation in $\P_l^\top$. Given any length-$N$ vector $\mathbf{x}$, the LLWA of $\mathbf{x}$ is defined as the length-$N$ vector $\mathbf{x}^\ddagger$ whose coefficient $\mathbf{x}^\ddagger[i]$ at node $i$ is a weighted convex combination of coefficients in the 27-neighborhood of $i$, $\mathcal{M}(i) = \{j | \mathbf{k}_{ij}=(\mathbf{n}_i - \mathbf{n}_j) \in \{-1, 0, 1\}^3 \}$:
\begin{equation}
\mathbf{x}^{\ddagger}[i] 
= \frac{\sum_{j \in \mathcal{M}(i)} w_{\mathbf{k}_{ij}} * \mathbf{x}[i] }
    {\sum_{j \in \mathcal{M}(i)} w_{\mathbf{k}_{ij}}} .
\label{eq:locally_linear}
\vspace*{-1ex}
\end{equation}
In vector form,
\begin{equation}
    \mbox{\it LLWA}(\mathbf{x}) = \mathbf{D}^{-1} \mathbf{W} \mathbf{x},
    \label{eq:matrix_LLWA}
\end{equation}
where $\mathbf{W} = [ w_{\mathbf{k}_{ij}}]_{N\times N}$ is the matrix of weights, and $\mathbf{D}$ is the diagonal degree matrix with diagonal elements $\mathbf{D}_{ii} = \sum_j w_{\mathbf{k}_{ij}} $.  Since the weights are space-invariant, the operation can be parametrized by 27 parameters, which may be trained.
% , which will become trainable parameters in the training process later on. 

To use LLWA in the prediction, we reconstruct the level-$L$ coefficients $F_L=\A_{L-1}^\top\cdots\A_l^\top F_l^*$ from the level $l$ coefficients $F_l^*$, apply LLWA to obtain $F_L^\ddagger=\mbox{\it LLWA}(F_L)$,
% We then apply this enhancement operation to the reconstructed attributes, namely $f^*_l = \bPhi_{l}F_{l}^{*}$ 
and then project the result back to scale $l+1$ using the normal equation \eqref{eq:normal_equation_F},
\begin{eqnarray}
    \lefteqn{\P_l^\top F_l^*
     = (\bPhi_{l+1}^{\top} \bPhi_{l+1})^{-1} \bPhi_{l+1}^{\top} \bPhi_L \mbox{\it LLWA}(F_L)} \\
    & = (\bPhi_{l+1}^{\top} \bPhi_{l+1})^{-1} \A_{l+1}\cdots\A_{L-1} \D^{-1}\W\A_{L-1}^\top\cdots\A_l^\top F_l^* .
    \label{eq:LLWA_pred_module}
\end{eqnarray}

\begin{figure}[t]
    \centering
    \includegraphics[height=0.1\textheight]{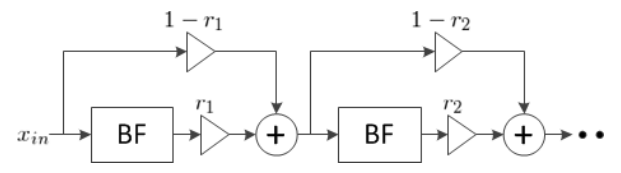}
    \vspace{-0.05in}
    \caption{Polynomial of bilateral filter}
    \label{fig:PolynomialBF}
\end{figure}
\begin{figure}[t]
    \centering
    % trim left lower right upper
    \includegraphics[%width=0.7\linewidth,
    height=0.06\textheight, trim=0.5in 4.75in 5.0in 1.125in,clip]{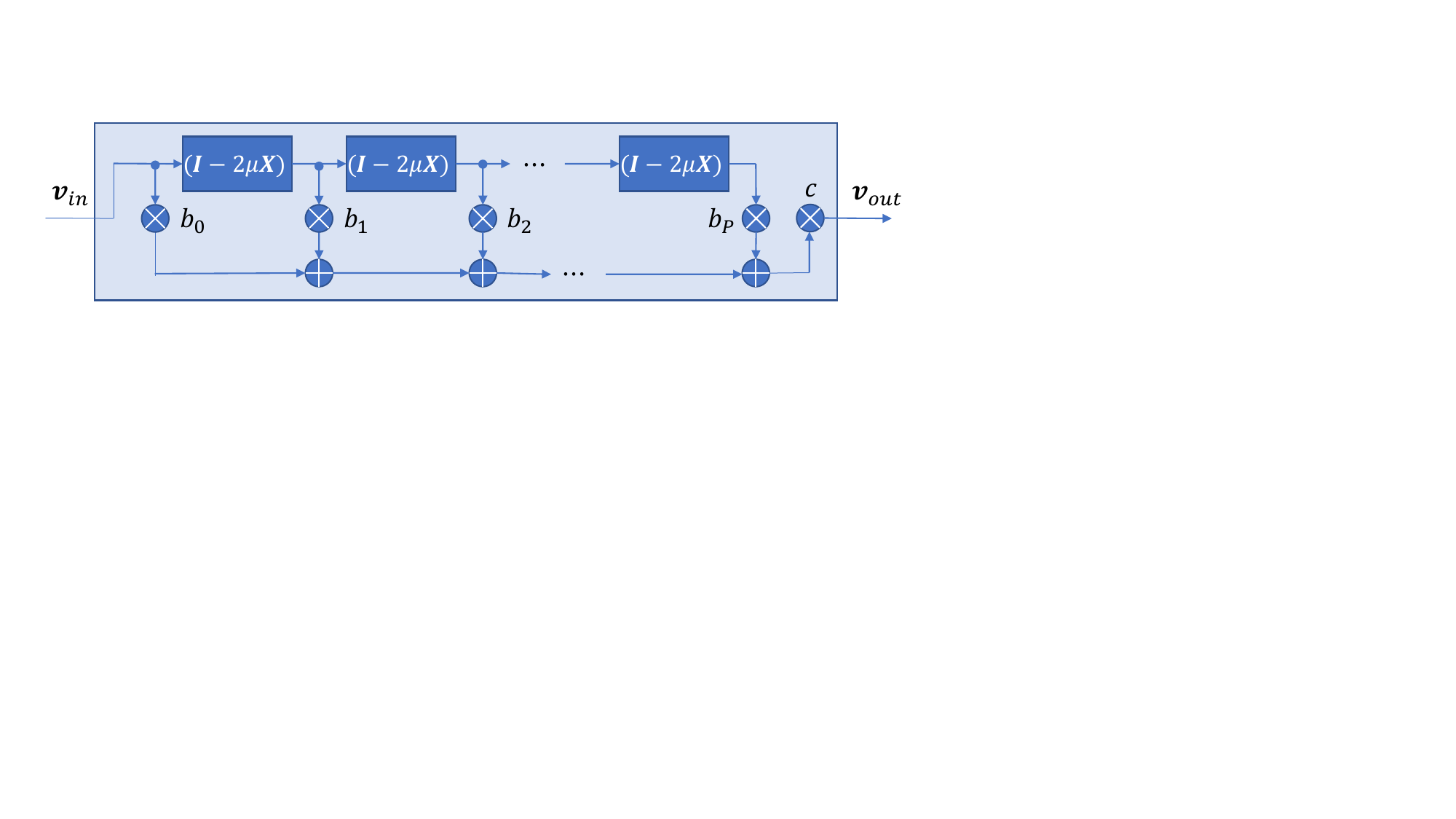}
    \caption{Multilayer feedforward network implementing the subnetworks $(\bPhi_l^{\top}\bPhi_l)^{-1}$ and $(\bPhi_l^{\top}\bPhi_l)^{-1/2}$ in Figs.\ \ref{fig:feedforward_network_block_diagram}. Here, $b_0,\ldots,b_P,c$ are coefficients of the $P$th order Taylor expansion of $x^{-1}$ and $x^{-1/2}$ around $1/(2\mu)$, where $\mu$ is an upper bound on the eigenvalues of $\mathbf{X}$.}
    \label{fig:taylor_expansion_block_diagram}
    \vspace*{-3ex}
% \caption{Neutral network architecture diagram}
\end{figure}
\subsection{Polynomial Bilateral Filter Prediction}

Next, we introduce another enhancement operation from \cite{antonio2013polybf}, the \textbf{polynomial bilateral filter} (PBF). The PBF is a polynomial version of the {\em bilateral filter} (BF).  As is well known, the BF is defined as in \eqref{eq:locally_linear}, when the weights in the the 27-neighborhood $\mathcal{M}(i)$ of node $i$ are given by
% We first calculate the filter weights from the inputs $\mathbf{x}$ using the 27-neighborhood of node i, $\mathcal{M}(i)$,
\begin{equation}
w_{\mathbf{k}_{ij}} = \exp \left( -\frac{\| \mathbf{n}_i-\mathbf{n}_j \|^2}{2\sigma_x^2} \right) \, \exp \left( -\frac{\| \mathbf{x}[i]-\mathbf{x}[j] \|^2}{2\sigma_y^2} \right) ,
\end{equation}
where $\sigma_x$ and $\sigma_y$ are the filter parameters, which may be trained.
In matrix form,
% Equation \eqref{eq:locally_linear} can be represented in matrix forms, with the bilateral filter weights $\mathbf{W} = [ w_{\mathbf{k}_{ij}}]_{N\times N}$ and $\mathbf{D}$ be the diagonal degree matrix where each diagonal element $\mathbf{D}_{ii} = \sum_j w_{\mathbf{k}_{ij}} $
\begin{equation}
    \mbox{\it BF}(\mathbf{x}) = \mathbf{D}_\mathbf{x}^{-1} \mathbf{W}_\mathbf{x} \mathbf{x} .
    \label{eq:matrix_BF}
\end{equation}
Here we have subscripted the weight and degree matrices to make explicit their dependence on the signal $\mathbf{x}$.  This dependence makes the BF both space-varying and
% Because of this dependence, the BF is
nonlinear in $\mathbf{x}$, though because of its form, it is often termed {\em pseudo-linear}.

To obtain the PBF, first we define $\mathbf{L}_\mathbf{x} = \mathbf{D}_\mathbf{x} - \mathbf{W}_\mathbf{x}$ as the combinatorial Laplacian matrix for the 27-neighborhood graph. We then interpret the bilateral filter \eqref{eq:matrix_BF} as a graph filter,
\begin{equation}
    \mbox{\it BF}(\mathbf{x}) = (\mathbf{I} - \boldsymbol{\mathcal{L}}_r) \mathbf{x} ,
    % \label{eq:matrix_BF}
\end{equation}
where $\boldsymbol{\mathcal{L}}_r = \mathbf{D}_\mathbf{x}^{-1} \mathbf{L}_\mathbf{x}$ is the random walk Laplacian matrix. The bilateral filter can then be generalized into a graph filter with spectral response function belonging to the class of polynomial functions with maximum degree $K$,
\begin{align}
    \textit{h}(\boldsymbol{\mathcal{L}}_r) &= r_0 \prod_{k=0}^{K} (\mathbf{I} - r_k \boldsymbol{\mathcal{L}}_r) , \\
    \mbox{\it PBF}(\mathbf{x}) &= \textit{h}(\boldsymbol{\mathcal{L}}_r) \mathbf{x} .
    \label{eq:polynomials_BF}
\end{align}
The polynomials coefficients $\{ r_k\}_{k=0}^K$ may be trained. As proved in \cite[Thm.~5.1]{antonio2013polybf}, operation $\textit{h}(\boldsymbol{\mathcal{L}}_r)$ can be implemented as a cascade of bilateral filters,
\begin{equation}
    (\mathbf{I} - r_k \boldsymbol{\mathcal{L}}_r) = (1-r_k)\mathbf{I}+r_k \mathbf{D}_\mathbf{x}^{-1} \mathbf{W}_\mathbf{x} .
\end{equation}
% as in Figure\;\ref{fig:PolynomialBF}. 
Thus, by replacing the \textit{LLWA} enhancement component with \textit{PBF}, we propose the following prediction module:
\begin{eqnarray}
    \lefteqn{\P_l^\top F_l^*
     = (\bPhi_{l+1}^{\top} \bPhi_{l+1})^{-1} \bPhi_{l+1}^{\top} \bPhi_L \mbox{\it PBF}(F_L)} \\
    & = (\bPhi_{l+1}^{\top} \bPhi_{l+1})^{-1} \A_{l+1}\cdots\A_{L-1} h(\mathcal{L}_r)\A_{L-1}^\top\cdots\A_l^\top F_l^* .
    \label{eq:GBF_pred_module}
\end{eqnarray}

\subsection{Feed-forward network \& training}
We implement the coding framework in the feed-forward network shown in Fig.~\ref{fig:feedforward_network_block_diagram}.  The encoding network consists of both blue and green components, while the decoding network consists of only green components.  The encoding network takes as input, at the top left, the point cloud attributes $\tilde{F}_L^*=[\mathbf{y}_i]$ represented as a $N_L\times r$ tensor, where $N_L$ is the number of points in the point cloud and $r$ is the number of attribute channels.

The operator $\bPhi_l^\top \bPhi_l$ is implemented by creating a graph of $N_{l}$ nodes, whose edge weights between nodes $i$ and $j$ are calculated using \eqref{eqn:PhiTPhi}.  Applying the operator to any length-$N_{l}$ vector involves only broadcasting the input to the edges $ij$ and aggregating by summing at nodes $j$. Similarly, the operator $\A_{l}$ is also implemented as a bipartite graph connecting $N_{l}$ nodes to $N_{l + 1}$ nodes.

Given a set $\Theta$ of trainable parameters, our feed-forward network model minimizes the reconstruction mean square errors, $D(\Theta) = \frac{1}{N_L} \| F^*_L - \hat{F}^*_L \|_2^2$, subject to a constraint on the bit rate,  $R(\Theta) \leq R_0$. This can be accomplished by minimizing the Lagrangian $J(\Theta) = D(\Theta) + \lambda R(\Theta)$, where the Lagrange multiplier $\lambda$ is chosen to attain the bit rate $R_0$. For training, we use the ``straight-through'' quantizer proxy $Q(\mathbf{x})=\mathbf{x}$ \cite{BalleCMSJAHT:21} and the arithmetic coding proxy
\vspace{-1ex}
\begin{equation}
    R(\{Y_l\}_{l=l_0}^{L}, \Theta) = -\frac{1}{N_L} \sum_{l=l_0}^L \text{log}_2 p(Y_l; m_l, b_l, \Delta),
    \vspace{-2ex}
\end{equation}
where $\{Y_l\}_{l=l_0}^{L}$ are all the coefficients at each level, namely $Y_{l_0} = \bar{F}_{l_0}^*$ and $Y_l = \overline{\delta F}_{l}^{\ddagger}$ for $l > l_0$, and 
% \vspace{-1ex}
\begin{align}
    &p(Y_l; m_l, b_l, \Delta) \\
    &= \textit{CDF}_{m_l, b_l}\left (Y_l + \frac{\Delta}{2} \right) - \textit{CDF}_{m_l, b_l}\left (Y_l - \frac{\Delta}{2} \right).
    \vspace{-2ex}
\end{align}
The $\textit{CDF}_{m_l, b_l}$ is modeled by the Laplace distribution at each level $l$ with scalar location $m_l$, scalar diversity $b_l$ and a uniform quantization step size $\Delta$ as trainable parameters.

\begin{figure*}[t]
    \begin{subfigure}{0.24\textwidth}
        \centering
        \includegraphics[width=\linewidth,trim=0.05in 0.0in 0.05in 0.0in,clip]{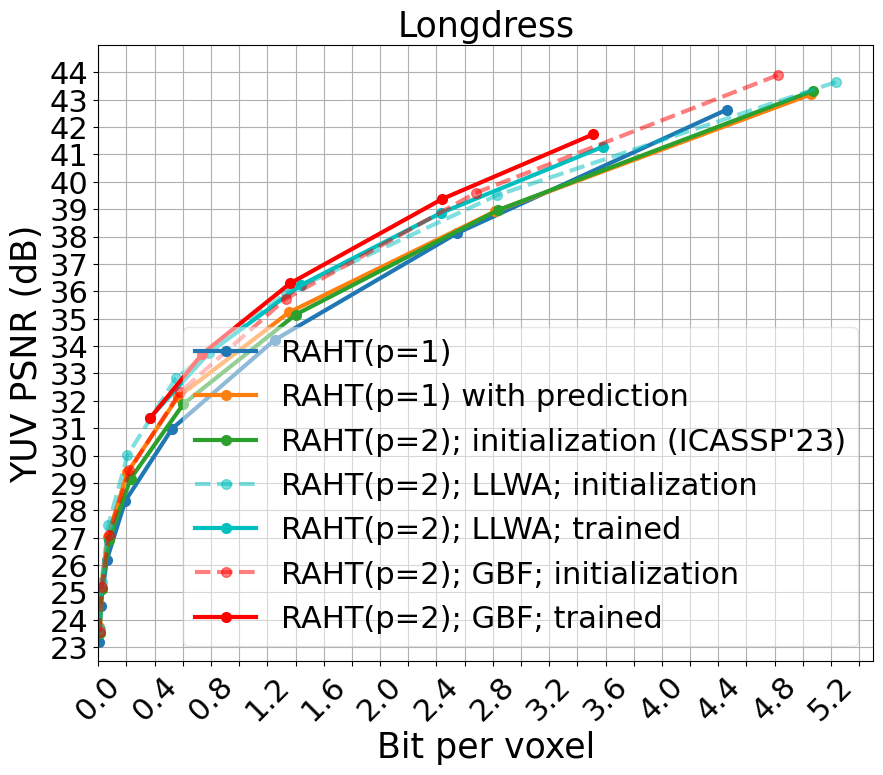}
        \caption{}
    \end{subfigure}
    \begin{subfigure}{0.24\textwidth}
        \centering
        \includegraphics[width=\linewidth,trim=0.05in 0.0in 0.05in 0.0in,clip]{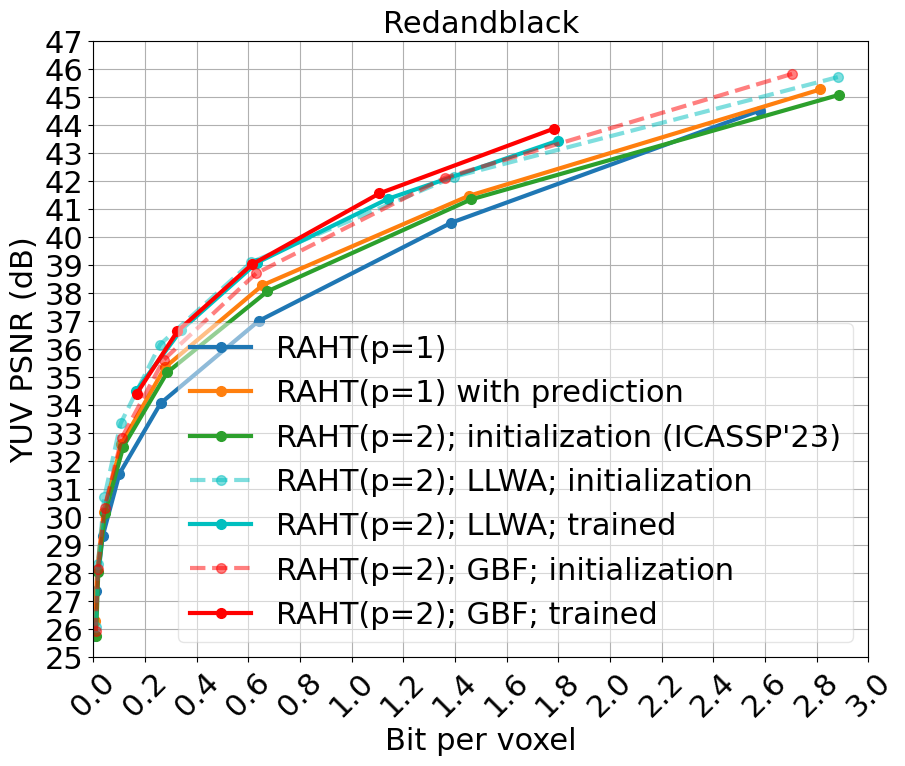}
        \caption{}
    \end{subfigure}
    \begin{subfigure}{0.24\textwidth}
        \centering
        \includegraphics[width=\linewidth,trim=0.05in 0.0in 0.05in 0.0in,clip]{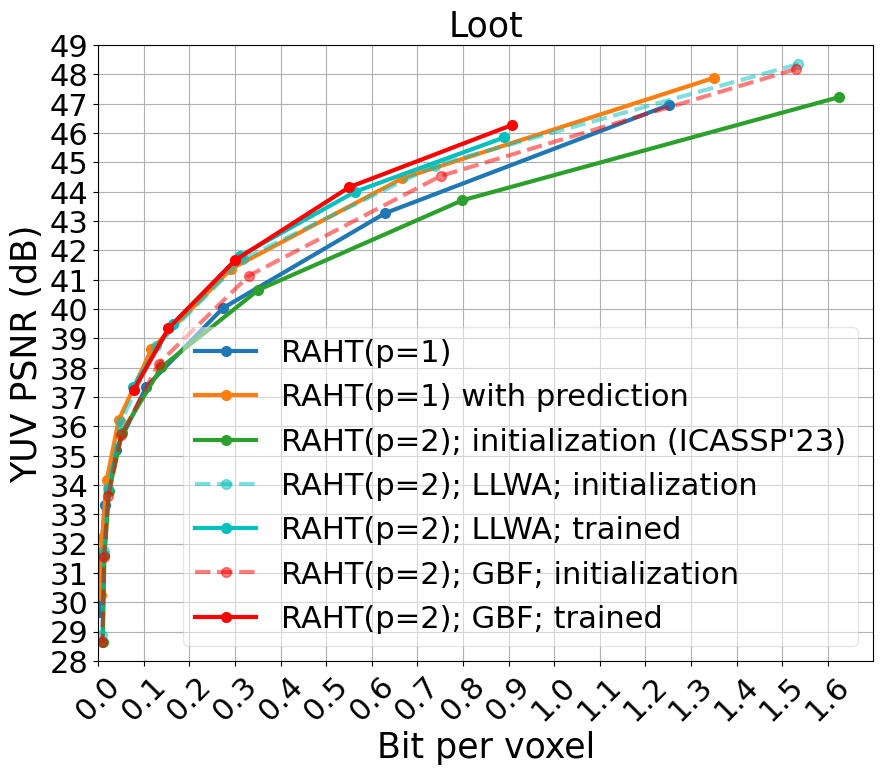}
        \caption{}
    \end{subfigure}
    \begin{subfigure}{0.24\textwidth}
        \centering
        \includegraphics[width=\linewidth,trim=0.05in 0.0in 0.05in 0.0in,clip]{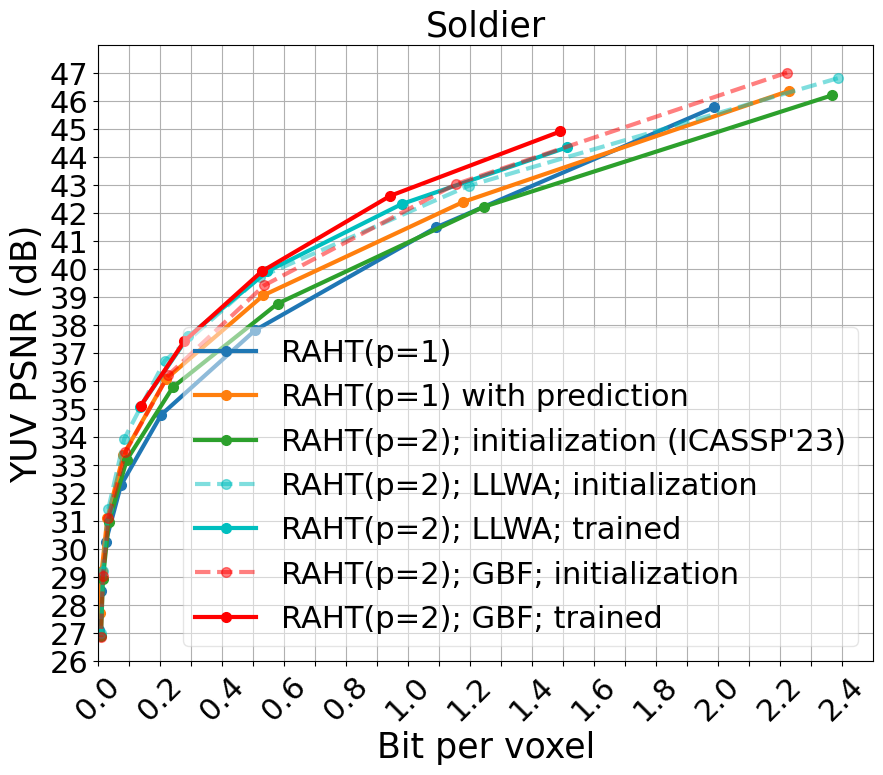}
        \caption{}
    \end{subfigure}
    \vspace{-0.05in}
    \caption{Rate-Distortion  curves: (a) \textit{Longdress}, (b) \textit{Redandblack}, (c) \textit{Loot}, (d) \textit{Soldier}}
    \label{fig:code_gain_performance}
\vspace*{-2ex}
\end{figure*}

Additional trainable parameters at each level $l$ include 3 kernels of size $3\times 3\times 3$ to parameterize $\mathbf{A}_l$, $\mathbf{A}_l^\top$, and $\mathbf{\Gamma}_l$, respectively. There are also 3 sets of $P+1$ polynomial coefficients to parameterize the sub-networks $(\bPhi_l^\top\bPhi_l)^{-1}$, $(\bPhi_l^\top\bPhi_l)^{1/2}$, and $(\bPhi_l^\top\bPhi_l)^{-1/2}$, respectively. Finally, the predictors have trainable parameters: $3\times 3\times 3$ kernel parameters for LLWA, and $\sigma_x$, $\sigma_y$ and $(K+1)$ polynomials coefficients $\{ r_k\}_{k=0}^K$ for PBF.

\vspace*{-0.5ex}
\section{Experiments}
\label{sec:results}
% \begin{figure*}[t]
%     \begin{subfigure}{0.24\textwidth}
%         \centering
%         \includegraphics[width=\linewidth,trim=0.05in 0.0in 0.05in 0.0in,clip]{ICASSP24/Figures/Result ICAPPS24 - Longdress.png}
%         \caption{}
%     \end{subfigure}
%     \begin{subfigure}{0.24\textwidth}
%         \centering
%         \includegraphics[width=\linewidth,trim=0.05in 0.0in 0.05in 0.0in,clip]{ICASSP24/Figures/Result ICAPPS24 - Redandblack.png}
%         \caption{}
%     \end{subfigure}
%     \begin{subfigure}{0.24\textwidth}
%         \centering
%         \includegraphics[width=\linewidth,trim=0.05in 0.0in 0.05in 0.0in,clip]{ICASSP24/Figures/Result ICAPPS24 - Loot.png}
%         \caption{}
%     \end{subfigure}
%     \begin{subfigure}{0.24\textwidth}
%         \centering
%         \includegraphics[width=\linewidth,trim=0.05in 0.0in 0.05in 0.0in,clip]{ICASSP24/Figures/Result ICAPPS24 - Soldier.png}
%         \caption{}
%     \end{subfigure}
%     \caption{Rate-Distortion  curves: (a)\textit{Longdress}, (b)\textit{Redandblack}, (c)\textit{Loot}, (d)\textit{Soldier}}
%     \label{fig:code_gain_performance}
% \end{figure*}

\subsection{Experiments Setup} \label{sec:experiment_setup}

Our training dataset comprises voxelized point clouds derived from 3D models (meshes) publicly available in sketchfab\footnote{https://sketchfab.com/blogs/community/sketchfab-launches-public-domain-dedication-for-3d-cultural-heritage/}. We picked 10 meshes that resemble the human form with various color textures to create the dataset. We densely sampled points on the faces of the meshes, calculated the points' colors from the provided UV texture map, and voxelized the points by quantizing the point positions into voxels and averaging the point colors in each voxel. 
Our evaluation dataset comprises the MPEG point clouds \textit{Longdress}, \textit{Redandblack}, \textit{Loot}, and \textit{Soldier} \cite{dEonHMC:2017}.
All point clouds have 10-bit resolution, so their octrees have depth of $L=10$ and $N_L \sim  1e^6$.

We implemented our model in Python using Jax. In all experiments, we picked $l_0=4$, $P=100$, $K=20$ and trained the model parameters described in Sec.~\ref{sec:implement}. Since we only have 6 levels from $l_0=4$ to $L=10$, instead training on whole 10-bit resolution point clouds per batch, we train on 6-bit ($64\times64\times64$) crops centered on random points. We use the Adam optimizer with a learning rate of 0.0001 in all configurations.

For evaluation, we use the adaptive Run-Length Golomb-Rice (RLGR) entropy coder \cite{Malvar:2006} to calculate a realizable bit rate. 

\subsection{Baselines}

As a baseline, we estimate the performance of the transform and predictor in the MPEG geometry-based point cloud codec (G-PCC) \cite{SchwarzEtAl:18, Wang2023predGPCC, GPCC:21} by using RAHT($p=1$), which is the core transform of MPEG G-PCC, coupled with a predictor that resembles the predictor in MPEG G-PCC,
\vspace{-0.13cm}
\begin{eqnarray}
     \P_l^\top F_{l}^*=  \textit{LLWA} (\mathbf{A}_l^\top F_{l}^{*}) ,
     \vspace{-0.13cm}
\end{eqnarray}
where the $3\times3\times3$ weights for LLWA are calculated as $w_{\mathbf{k}_{ij}}=exp(-\frac{\| \mathbf{k}_{ij} \|^2}{2*(0.5)^2})$ with $\mathbf{k}_{ij} \in \{ -1, 0, 1 \}^3$.
% With RAHT($p=1$), the matrix $\bPhi_l^\top\bPhi_l$ is just a diagonal matrix \cite{ChouKK:20} so the operators $(\bPhi_l^\top\bPhi_l)^{-1}$ and $(\bPhi_l^\top\bPhi_l)^{-\frac{1}{2}}$ are calculated exactly.

We also show the results in our previous work \cite{do2023volumetric} to demonstrate the effectiveness of the prediction module and the training process.

\subsection{Results}
For our models, we first initialize the weights of $\A_l=[a_{ij}^l]$ such that they correspond to volumetric B-splines of order $p=2$, namely RAHT($p=2$). For the prediction, we test the two different predictors introduced in Sec.~\ref{sec:implement}: \textit{LLWA} and
\textit{PBF}. Since we can fully interpret all components in our models, we are able to initialize the model parameters to attain reasonable performance before any training begins, and then we are able to improve the performance by training according to Sec.~\ref{sec:experiment_setup}.

In Fig.~\ref{fig:code_gain_performance}, we show YUV PSNR as a function of bits per occupied voxel. Here, we use the RLGR entropy coder \cite{Malvar:2006} to code the coefficients of each color component (Y, U, V) separately. Overall, comparing with the baseline of RAHT($p=1$) with prediction, the plot shows that our models improve by more than 1 dB in PSNR over a wide range of bit rates, or alternatively reduce the bit rate by 17-23\%, with either LLWA or PBF predictors (before or after training). 

Our coding improvements over the baseline model are largely attributable to the higher B-spline order, RAHT($p=2$), coupled with prediction. This is expected because the representations at scale $l$ are the \textit{piecewise trilinear} functions of RAHT($p=2$), which contain much more detail than the \textit{piecewise constant} functions of RAHT($p=1$).  The former can be used for better prediction of the next scale $l+1$. Fig.\ref{fig:reconstructed_PCs} shows that blocking artifacts are much less visible with our model, especially at boundaries between objects, where there are continuous transitions of colors.

 The additional prediction components in our framework also significantly improve over previous works.  A nearly 45\% bit rate reduction is shown by comparing the RD curves that have prediction modules with actual RAHT($p=1$) or non-critical RAHT($p=2$) introduced in \cite{do2023volumetric}.
 
 Furthermore, training also improves performance, as shown by comparing each model after training with itself before training (dashed curves with same color). Hence, the structure of our feed-forward network is not only fully understandable but it can also be trained to adapt to arbitrary point clouds data.

\vspace*{-0.05in}
\section{Conclusion}
\label{sec:conclude}
We extend previous works, including MPEG G-PCC, that use a volumetric approach for 3D point cloud attribute coding, by proposing a polynomial bilateral filter for predicting the upsampled signals, and training.  Experiments show that our learned feedforward network outperforms previous schemes, including a G-PCC-like scheme, by up to $1$-$2$ dB in PSNR.

%\vfill
%\pagebreak
%KK%
\vfill
\pagebreak
%KK%

\bibliographystyle{IEEEbib}
\scriptsize
\bibliography{ref}

\end{document}